\documentclass[epj]{svjour}
\usepackage{graphics}
\begin{document}
\authorrunning{E.~Geraci et al.,}
\titlerunning{Isotope analysis at intermediate energies}
\title{Isotope analysis in central heavy ion collisions at intermediate energies}
\author{NUCL-EX Collaboration: E.~Geraci\inst{1}, U.~Abbondanno\inst{2}, L.~Bardelli\inst{3}, 
S.~Barlini\inst{4}, M.~Bini\inst{3}, M.~Bruno\inst{1}, F.~Cannata\inst{1}, 
G.~Casini\inst{3}, M.~Chiari\inst{3}, M.~D'Agostino\inst{1}, 
J.~De~Sanctis\inst{1}, 
A.~Giussani\inst{5}, F.~Gramegna\inst{4}, V.~L.~Kravchuk\inst{4}, 
A.~L.~Lanchais\inst{4}, 
P.~Marini\inst{1}, A.~Moroni\inst{5}, A.~Nannini\inst{3},
A.~Olmi\inst{3}, 
A.~Ordine\inst{6}, 
G.~Pasquali\inst{3}, S.~Piantelli\inst{3}, G.~Poggi\inst{3}, 
and G.~Vannini\inst{1}
}                     
%
%
\institute{Dipartimento di Fisica and Sezione INFN di Bologna \and
 Dipartimento di Fisica and Sezione INFN 
di Trieste \and Dipartimento di Fisica and Sezione INFN di Firenze 
\and INFN, Laboratori Nazionali di Legnaro \and 
Dipartimento di Fisica and Sezione INFN di Milano \and Dipartimento di
Fisica and Sezione 
INFN di Napoli}
\date{Received: date / Revised version: date}
%
\abstract{
Symmetry energy is a key quantity in the study of the equation of state of asymmetric nuclear 
matter. Heavy ion collisions at low and intermediate energies, performed at Laboratori Nazionali 
di Legnaro and Laboratori Nazionali del Sud, can be used to extract information on the symmetry 
energy coefficient $C_{sym}$, which is currently poorly known but relevant both for astrophysics and 
for structure of exotic nuclei.
}
\maketitle
Heavy-ion collisions can be considered an excellent tool to explore the nuclear equation of 
state (EOS) of nuclear matter in laboratory controlled conditions. With the availability of 
radioactive beam facilities, the isospin is being extensively explored. One of the goals of 
these studies is to provide a better knowledge of the symmetry term of
the EOS. In par\-ti\-cu\-lar, stable and radioactive beams over a wide range of $N/Z$ asymmetries 
allow to explore the asymmetric nuclear EOS and the density dependence 
of the symmetry energy.

Symmetry energy and its density dependence determine several properties 
of neutron stars as well as features (binding energy and rms radii) of 
exotic nuclear systems as neutron halo nuclei.

In statistical and dynamical models, the isotopic composition of fragments emitted in 
multi-fragmentation phenomena, in central heavy-ion collisions at intermediate energies, 
are sensitive to the density dependence of the symmetry term and therefore it can provide 
information on symmetry energy at low density. Indeed, in these reactions complex fragments 
are expected to be formed at low densities ($\rho \sim 0.1-0.5\rho_0$) and temperatures T=3-5 MeV.
The study of the production yields of isotopically resolved nuclear
particles and fragments 
can complete the knowledge of the EOS, as a function of N/Z, and is
essential in searching for 
a possible occurrence of critical phenomena associated to fluctuations in the proton concentration of 
asymmetric nuclear matter.
In the grand canonical model, the ratio of the primary fragment yield (for a given isotope 
emitted in two different reactions differing only in N/Z ratio) depends exponentially on the 
neutron and proton number, the so called isoscaling \cite{1a}, by the
relation:
 $$R_{21}=Y_1(N,Z)/Y_2(N,Z)=Cexp(\alpha N+\beta Z)$$
 where $Y_1$ and $Y_2$ are the yield of a given isotope respectively in
 the neutron rich and neutron deficient system, C is an
 overall constant and $\alpha$ and $\beta$ are the isoscaling parameters.
 It turns out that~\cite{1} the isoscaling parameter
$\alpha$ is almost independent on the secondary de-excitation of the primary fragments inducing 
to use it as a robust observable to extract information on the symmetry energy. Indeed, 
different statistical and dynamical models~\cite{2} relate $\alpha$ to the symmetry energy coefficient 
$C_{sym}$ via
$$\alpha=\frac{4 C_{sym}}{T}(\frac{Z_1^2}{A_1^2}-\frac{Z_2^2}{A_2^2})$$
where T, Z and A are the temperature, charge and mass of the fragmenting system. 
In this way an estimate of the symmetry energy coefficient can be obtained whenever 
the isoscaling is observed and the temperature and the Z/A ratio for the fragmenting 
systems are determined.
\begin{figure} [ht]
\resizebox{1.\columnwidth}{!}{\includegraphics{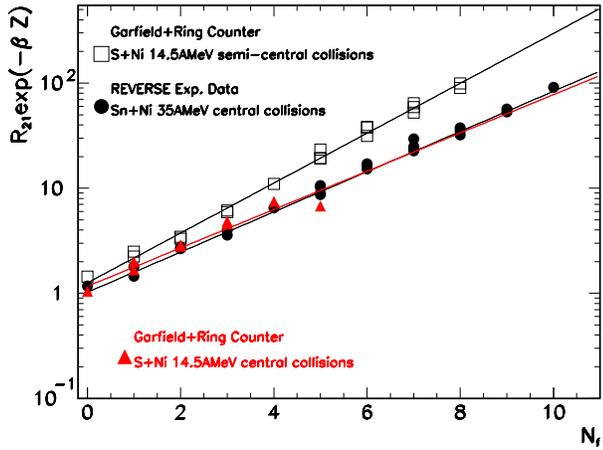}}
\caption{
Scaled isotopic ratio as a function of the neutron number of fragments
for Sn+Ni reactions at 35 AMeV and S+Ni  reactions at 14.5 AMeV.}
\label{fig1}       
\vspace{-0.25cm}
\end{figure}

We present here two different sets of data collected with 4 $\pi$ detectors at Laboratori Nazionali del 
Sud (LNS) and Laboratori Nazionali di Legnaro (LNL). In particular $^{124}$Sn+$^{64}$Ni and
 $^{112}$Sn+$^{58}$Ni 
reactions at 35 AMeV incident energy were studied by using the 688 Si-CsI telescopes of the 
forward part ($1^o \le \theta_{lab} \le 30^o$) of CHIMERA multi-detector at LNS, in the framework of the REVERSE 
experiment~\cite{3}. 
The most central collisions were selected by means of a multidimensional analysis
of the experimental observables. A detailed analysis of the yields of the detected isotopes of 
light fragments ($3 \le Z \le 8$) provided information on breakup temperatures of the emitting sources,
by means of the double isotope ratio thermometers, which resulted $\sim$ 4 MeV. Isoscaling has been observed for 
fragments ($1 \le Z \le 8$), with $\alpha$ equal to 0.44.

In Fig.~\ref{fig1} the scaled ratio
$$S(N_f)=R_{21}exp(-\beta Z)=Cexp(\alpha N)$$
is plotted as a function of fragment neutron number.
For the Sn+Ni reactions at 35 AMeV an estimate of the symmetry energy coefficient can be 
obtained considering the Z/A ratio suggested by SMM (Statistical Multifragmentation Model), 
which was used only to reconstruct (backtracing) the characteristics
 of the sources formed in
 central collisions. A $C_{sym}$ close to 
 12 MeV has been evaluated, in agreement with similar estimates~\cite{4}
 de\-mon\-stra\-ting that $C_{sym}$ for
 hot light nuclei in multifragmentation decreases with the excitation energies from 
 25 MeV for very peripheral collisions to 15 MeV or lower for central collisions.
Isotope analysis has been performed also for fragments produced in central and semi-central
collisions at lower incident energy. Beams of $^{32}$S were accelerated at 14.5 AMeV on 
$^{58}$Ni and $^{64}$Ni targets at LNL.
 The rea\-ction products were detected with the forward GARFIELD 
chamber~\cite{5} coupled to the Ring Counter (RCo)~\cite{6} , a
forward-angle apparatus specially designed for small polar angles.
 The very good mass resolution of 
the RCo Si-CsI telescopes and the good coverage of the phase space of the whole 
apparatus enable us to carry out accurate thermodynamics of excited systems with 
different N/Z at low energies  where the hot 
composite systems is expected to enter the liquid-gas coexistence~\cite{7}.
 Central collisions were selected imposing the presence of at least three 
 fragments in Garfield and RCo apparata and a flat distribution of the
 $cos(\theta_{flow})$, corresponding to $ \theta_{flow} \ge 60^o$.
 Events characterized by $\theta_{flow} \le 30^o$ and a total fragment
 multiplicity greater or equal 3 were provisionally labelled as semi-central.
 A deeper analysis on these latter events is in progress in order to evaluate
 possible contaminations of semi-central events from residues coming
 from deep-inelastic collisions.
 The detected isotopes of light 
 fragments ($1 \le Z \le 8$) have been used to extract double isotope ratio temperature of the 
 formed sources, resulting respectively equal to 3.5 and 3.2 MeV. The isoscaling analysis 
 has been performed and the scaled ratio is displayed in Fig.~\ref{fig1} for both reactions.
  The parameter $\alpha$ is influenced both by the difference in (N/Z) of the two systems 
  used to construct $R_{21}$ and by the excitation energy of the fragmenting system.
   The S+Ni central collisions show a slope lower than the S+Ni semi-central 
   reactions, in so far as increasing the excitation energy the $\alpha$
   parameter decreases. Moreover, the S+Ni central collisions exhibit
  an $\alpha$ parameter very close to the Sn+Ni central collisions
  one.
  This is due to the larger difference in the N/Z ratio of the
  two systems used to construct $R_{21}$ for the Sn+Ni reaction, inducing a reduction of the $\alpha$
  parameter. 
  
 In order to extract the symmetry energy for 
   these reactions an estimate of the Z/A ratio of the fragmenting systems is necessary.
    To this end a comparison of experimental quantities with dynamical and 
    statistical model has been undertaken. As a preliminary attempt one can estimate the Z/A ratio 
    in the hypothesis that it does not change from the entrance channel value.
     In this case one obtains $C_{sym}=12$ MeV for central collisions and
     $C_{sym}=14$ MeV for 
     semi-central collisions, respectively. Even if this evaluation sensitively 
     depends on the Z/A of the sources,which has to be refined by
   model predictions, the trend of Csym is consistent with other 
     studies that reveal a $C_{sym}$ closer to the standard value (25 MeV for normal 
     nuclei at saturation density) at lower excitation energies (E*/A $\sim 2$ MeV)~\cite{8}, decreasing 
     monotonically as the excitation energy E*/A grows up to 5-8MeV~\cite{4}.

\end{document}